\documentclass[10pt,letterpaper]{article}
\usepackage[top=0.85in,left=2.75in,footskip=0.75in]{geometry}

\usepackage{amsmath,amssymb}

\usepackage{changepage}

\usepackage[utf8x]{inputenc}

\usepackage[warn]{textcomp}
\usepackage{marvosym}

\usepackage{cite}

\usepackage{nameref,hyperref}

\usepackage{microtype}
\DisableLigatures[f]{encoding = *, family = * }

\usepackage[table]{xcolor}

\usepackage{array}

\newcolumntype{+}{!{\vrule width 2pt}}

\newlength\savedwidth

\raggedright
\usepackage[aboveskip=1pt,labelfont=bf,labelsep=period,justification=raggedright,singlelinecheck=off]{caption}

\makeatletter
\renewcommand{\@biblabel}[1]{\quad#1.}
\makeatother

\usepackage{lastpage,fancyhdr,graphicx}
\usepackage{epstopdf}
\fancyheadoffset[L]{2.25in}
\fancyfootoffset[L]{2.25in}
\usepackage{times}
\usepackage{epsfig}
\usepackage{graphicx}
\usepackage{amsmath}
\usepackage{amssymb}

\usepackage{makecell}
\usepackage{booktabs} %
\usepackage{supertabular}
\usepackage{tabularx}
\usepackage{ragged2e}

\definecolor{mwcolor}{rgb}{.6,0.4,0}

\def \numberOfBoxes {227,810 } 
\def \automaticnumberOfBoxes {96k }
\def \numberOfPaintings {19,325 } 
\def \numberOfMaterialInstances {123,244 }

\def \dbwebsite {\href{https://materialsinpaintings.tudelft.nl/}{\textcolor{red}{materialsinpaintings.tudelft.nl}}}

\newcommand{\fig}[1]{Fig.~\ref{#1}}
\newcommand{\tabl}[1]{Table~\ref{#1}}

\begin{document}
\vspace*{0.2in}

\begin{flushleft}
{\Large
\textbf\newline{Materials In Paintings (MIP): An interdisciplinary dataset
  for perception, art history, and computer vision} 
}
\newline
\\
Mitchell J.P. Van Zuijlen* \textsuperscript{1},
Hubert Lin \textsuperscript{2},
Kavita Bala \textsuperscript{2},
Sylvia C. Pont \textsuperscript{1},
Maarten W.A. Wijntjes \textsuperscript{1}

\bigskip
\textbf{1} Perceptual Intelligence Lab, Delft University of Technology, Delft, The Netherlands
\\
\textbf{2} Computer Science Department, Cornell University, Ithaca, New York, United States of America
\\
\bigskip

* Corresponding author: m.j.p.vanzuijlen@tudelft.nl

\end{flushleft}


\section*{Abstract}

A painter is free to modify how components of a natural scene are depicted, which can lead to a perceptually convincing image of the distal world.  
This signals a major difference between
photos and paintings: paintings are explicitly created for human
perception. Studying these painterly depictions could be beneficial to a multidisciplinary audience. 
In this paper, we capture and explore the painterly depictions of materials to enable the study of 
depiction and perception of materials through the artists’ eye. We annotated a dataset of 19k paintings
with 200k+ bounding boxes from which polygon segments were automatically extracted. Each bounding box was assigned a coarse label (e.g.,
fabric) and a fine-grained label (e.g., velvety, silky). We demonstrate the cross-disciplinary utility of our dataset by presenting novel 
findings across art history, human perception, and computer vision. Our experiments include analyzing the distribution of 
materials depicted in paintings, showing how painters create convincing depictions using a
stylized approach, and demonstrating how paintings can be used to build more robust computer vision models. 
We conclude that our dataset of painterly material depictions is a rich source for gaining insights into the depiction and perception of materials across multiple disciplines. 
The MIP dataset is freely accessible at \dbwebsite. 

\section*{Introduction}

\label{section:introduction}

Throughout art history, painters have invented numerous ways to
transform the three-dimensional world onto flat surfaces
\cite{panofsky2020perspective,white1957birth,kemp1990science,pirenne1970optics}.
These transformations can be described by the geometry of the
projection or in terms of two-dimensional drawing rules. Willats
\cite{willats1997art} noticed that for each projection there exists a
set of drawing rules, but not vice versa. This illustrates an
interesting and fundamental asymmetry that is characteristic to the visual
perception of pictures: a depiction does not necessarily have to
originate from a physically correct projection\cite{Cavanagh2005a}. On
one hand, this makes paintings unsuited as ecological stimulus
\cite{gibson1978ecological}. On the other hand, as Gibson
acknowledges, paintings are the result of endless visual
experimentation, and therefore, indispensable for the study of visual
perception.

In perception, \textit{distal} refers to the outside world, while the related concept \textit{proximal} refers to the projection of the distal world experienced on our senses i.e., the retinal image. The process of visual perception can be described as inferring properties of the distal stimulus from the available proximal information\cite{Anderson2011}. For example, perceived reflectance (lightness) is
a distal property that is deduced from light on the retina
(brightness) \cite{gilchrist2007lightness}, or a distal circle is
inferred from a proximal ellipse \cite{hammad2008ellipses}. This
is akin to `recovering intrinsic scene characteristics' (which
are distal) from images (which are proximal) in computer
vision~\cite{Barrow1978} .

A painter might work with `2D drawing rules', but the goal usually is not to create an optically corrected projection, rather a painter strives to create a perceptually correct depiction.  The artist does not copy a
retinal image \cite{perdreau2011artists} (which would make the painter
effectively a biological camera) but rather iteratively adapts
templates until they `fit' perceptual awareness \cite{GombrichE1960}. In essence, a painting depicts the perceived distal world through a proximal stimuli. 

The depiction and perception of pictorial space in paintings \cite{willats1997art, panofsky2020perspective,white1957birth,kemp1990science,pirenne1970optics} has received much 
more attention than the depiction and perception of materials. As with the depiction of space, a painter is not concerned whether a material depiction is optically or physically correct. Instead, a painting is explicitly designed for human viewing and is only intended to be perceptually convincing. Human observers are able to visually categorize and identify depicted materials and material properties \cite{Fleming2017}, despite a painting consisting only of paint and oils, similar to how we can perceive depth from a flat painting. Furthermore, for these painted materials, we can perceive distinct  material properties such as glossiness, softness, transparency, etc \cite{van2020painterly, DiCicco2019, di2020material}. A single material category (e.g., fabric) can display a large variety of these material properties, which demonstrates the enormous variation in visual appearance of materials. This variation in materials and material properties has received relatively little attention. In fact, the perceptual knowledge that is captured in the innumerable artworks throughout history can be thought of as the largest perceptual experiment in human history and it merits detailed exploration.
The starting point for such an exploration is the creation of datasets that relate artworks to material perception. In this study, we introduce an accessible collection of material depictions in paintings, with which we hope to facilitate  both perceptual, computational, and art historical research into materials.

\subsection*{A simple taxonomy of image datasets} 
In the current study we are primarily interested in the perception and depiction of materials and material properties \cite{Fleming2015, Adelson2001, Anderson2011, Fleming2017}. However, the use and creation of art-perception datasets is of broader interest. We  propose a simple taxonomy of three image dataset usages: 1) perceptual, 2) ecological and, 3) computer vision usage. We expand on each of these three usages below and contextualize our dataset within this taxonomy.

\paragraph*{Perceptual datasets.}

To understand the human visual system, stimuli from perceptual datasets can be  used in an attempt to relate the evoked perception to the visual input. We can roughly categorize three types of
stimuli used for visual perception: natural, synthetic and manipulated. The first represent `normal' 
images of objects, materials and scenes as they can be found in reality. Experimental design with such
stimuli often attempts to relate the evoked perceptions to natural image statistics within the images or physical characteristics of the contents captured in the images. 
Some examples of uses of natural stimuli datasets include the memorability of pictures in general
\cite{isola2011makes} or more specifically the memorability of faces \cite{bainbridge2013intrinsic}.
In another example, images of natural, but novel objects were used to understand what underlies the
visual classification of objects \cite{horst2016novel}. The second type,
synthetic stimuli, are created artificially. Synthetic stimuli  might
represent the real world, but often contain
image statistics that deviate greatly from natural image statistics. For example, \cite{borkin2013makes} used a set of synthetic stimuli to test for memorability of data visualizations. 
Both natural and synthetic images can be manipulated, which leads to
the third type of stimuli. Manipulated stimuli are often used to
investigate the effect of image manipulations by comparing them to the
original (natural or synthetic) image. Here the manipulations function
as the independent variables. For example
\cite{ohlschlager2017scegram} created a database of images that contain scene inconsistencies that can be used to study the compositional rules of our visual environment.  In
another example, a stimulus set consisting of original and texture
(i.e., manipulated) versions of animals found that perceived animal size is mediated by mid-level image statistics \cite{long2018mid}.

The advantage of using manipulated or synthetic images is that perceptual judgments can be compared 
to some independent variable, which is not available for natural images. Paintings are a special case: while 
they are being created they are a synthetic image that is rendered using oils and paints. However,
when finished they are also real, physical objects. While retrieving the veridical data is impossible 
for almost all paintings, the advantage of using paintings is that it can often be seen, or (historically)
inferred, how the painter created the illusory realism. Even if it can not be seen with the naked eye, chemical and physical analysis can be performed. In \cite{di2020material} a
perceptually convincing depiction of grapes was recreated using a 17th century, explicitly written-down recipe. In the reconstruction, the depiction was created one layer at a time, each representing a separate and perceptually diagnostic image feature of the grape. In this way, paintings can give access to proximal information. Therefore, studying paintings in addition to
more traditional stimuli like photos or renderings, can enrich our
understanding of human perception.  It should be noted that in this paper we focus on the image structure of the painting instead of the physical object. In other words, we focus on what is depicted within paintings and our data and analysis is limited to pictorial perception. In the remainder of this paper, when we mention \textit{paintings}, we mean \textit{images of paintings}.

Throughout history, painters have studied how to
trigger  the perceptual system and create convincing proximal depictions of complex distal properties of the world. This resulted in \textit{perceptual shortcuts}, i.e., stylized depictions of complex properties of the distal world  that trigger a robust perception. The steps and painterly techniques applied by a painter to create a perceptual shortcut can be thought of as a perception-based recipe. Following such a recipe results in a perceptual shortcut, which is a  depiction that gives the visual system the required inputs to trigger a perception. Many of the successful depictions are now available in museum collections. As such, the creation of art throughout history can be seen as one massive perceptual experiment. Studying perceptual shortcuts in art, and understanding the cues, i.e., features required to trigger perceptions, can give insights into the visual system. We will demonstrate this idea by analyzing highlights in paintings and photos. 

\paragraph{Ecological datasets.}

To understand how the human visual system works it is important to understand
what type of visual input is given by the environment.  Visual ecology encompasses all the visual input and can be subdivided into natural and cultural ecology. Natural ecology reflects all  which is found in the physical world. For example,
to understand color-vision and cone cell sensitivities it is relevant to know
the typical spectra of the environment. For this purpose, hyperspectral images
\cite{foster2006frequency,nascimento2016spatial} can be used, in this case to
investigate color metamers (perceptually identical colors that originate from
different spectra) and illumination variation.  In another example, a dataset
of calibrated color images were used to understand color constancy
\cite{ciurea2003large} (the ability to discount for chromatic changes in
illumination when inferring object color). The SYNS database was used to relate
image statistics to physical statistics
\cite{adams2016southampton}.  Another dataset contains photos taken in
Botswana \cite{tkavcik2011natural} in an area that supposedly reflects the
environment of the proto-human and was used to investigate the evolution of the
human visual system. Spatial statistics of today's human visual ecology are clearly
different from Botswana's bushes as most people live in urban areas that are shaped by humans.
For example, a dataset from \cite{olmos2004biologically} was used to compute the
distribution of spatial orientations of natural scenes \cite{girshick2011cardinal}.

It is important to note that the majority of paintings in our dataset are painterly representations of the physical world around us that only loosely
reflect the natural visual ecology, but strongly represent visual
cultural ecology. They have influenced how people see and depict the world and have
influenced visual conventions up to contemporary cinematography and
photography. The recent surge in publicly available digitized  art
works, combined  with the availability of advanced image analysis
algorithms such as neural networks, has lead to a new branch of Art
History: Digital Art History. Similar to the analysis of the human
physical environment, Digital Art History concerns itself for example with the
digitized analysis of artworks, artistic style \cite{Saleh2016}
and beauty \cite{de2015quantitative}, or local pattern similarities between artworks \cite{Shen_2019_CVPR}. 

\paragraph{Computer vision datasets.}

Today, the majority of image datasets originate from research in computer
vision.  One of the first relatively large datasets representing object
categories \cite{fei2004learning} has been used to both train and evaluate
various computational strategies to solve visual object recognition. The
ImageNet and CIFAR datasets \cite{imagenet, cifar} are regarded to be standard image recognition
datasets for the last decade of research ifn deep learning vision systems. Traditionally most visual research has been concerned with object
classification but recently material perception has received increasing attention
\cite{Adelson2001b, bell13opensurfaces, minc, caesar2018coco}.

Within vision science, paintings are considered as representing a special class of
images, which deviate from natural photographs, as they are explicitly designed for
viewing by humans \cite{Graham2010StatisticalPerception}. 
The visual difference introduced by painterly depiction does not pose any significant difficulties to the human visual
system, however it is more challenging for computer vision systems
as a result of the domain shift
\cite{patel2015visual,wang2018deep,wilson2020survey}. Differences between painting images and photographic datasets include for instance composition, textural properties, colors and tone mapping, perspective, and style. As for composition, photos
in image datasets are often `snapshots', taken with not too much thought given
to composition, and typically intended to quickly capture a scene or event. In
contrast, paintings are artistically composed, and prone to historical style
trends. Therefore,  photos often contain much more composition variation
relative to paintings. Within paintings, composition can vary greatly between different styles. 
The human visual system can \emph{distinguish} styles -- for example, Baroque vs. Impressionism -- and also implicitly judge whether two paintings are stylistically
similar. Research in style or artist classification, as well as neural networks that perform style
transfer, attempt to model these stylistic variations in art
\cite{saleh2015largescale,nstsurvey}. Humans can also \emph{discount} stylistic
differences, for example, identifying the same person or object depicted by
different artists.  Similarly, work in domain adaptation
\cite{patel2015visual,wang2018deep,wilson2020survey} focuses on understanding
objects or stuff across different image styles.

Depending on the end goal for a computer vision system, it can be important to
learn from paintings directly. When the end goal is to detect pedestrians for a
self-driving car, learning from real photos, videos, or renderings of
simulations can suffice. However, if the goal is to simulate general visual
intelligence, multi-domain training sets seem essential. Furthermore, if the goal is to create computer vision systems with a perception that matches human vision, training on paintings could be very beneficial. Paintings are explicitly created by and for human perception and therefor contain all the required cues to trigger robust perceptions. Therefor, networks trained on paintings are implicitly trained on these perceptual cues. 

\paragraph{The multifaceted nature of datasets.} 

While we have distinguished the broad purposes of datasets and exemplified each
with representative datasets, it is important to keep in mind that these
datasets can serve multiple goals across the taxonomy.  For example, the Flickr
Material Database \cite{Sharan2009} was initially created as a perceptual dataset to study how
quickly human participants were capable of recognizing natural materials.  However, since then it has also often been used as a computer vision dataset, including by the original authors themselves \cite{Sharan2013}. The dataset presented in this paper is explicitly designed with this
multidisciplinary nature in mind. 

\section*{Dataset collection and annotation}

Our dataset consists of 19K paintings with crowd-sourced bounding box
annotations over 15 material categories. We further distinguish these coarse
material categories into over 50 fine-grained categories. Finally, we
automatically extract polygon segments for each bounding box. The annotated
dataset will be made publicly available. All paintings, bounding
boxes, labels, and metadata  are available at \dbwebsite

The data collection was executed in multiple stages. Here we give an itemized
overview of each stage and subsequently we discuss each stage in depth. The
first two stages were conducted as part of a previous study
\cite{van2020painterly}, but we provide details here for completeness. Participants were recruited via Amazon Mechanical Turk (AMT). A total of 4451 unique AMT users participated in this study. Data collection was approved by the Human Research Committee of the Delft University of Technology and adheres to the Declaration of Helsinki.

\begin{enumerate}
    \item First, we collected a large set of paintings.
    \item  Next, human observers on the AMT platform identified which
coarse-grained materials they perceived to be present in each painting (e.g., ``is there wood depicted in this painting?"). 
    \item Then, for paintings identified to contain a specific
material, AMT users were tasked with creating a bounding box of that
material in that painting.
    \item Lastly, AMT users assigned a fine-grained material label to
bounding boxes (e.g., processed wood, natural wood, etc.).
\end{enumerate}

\subsection*{Collecting paintings} 
We collected \numberOfPaintings paintings from 9 online, open-access
art galleries. The details of these art galleries are presented in \tabl{galleries}. Images were downloaded from the online galleries, either using web scraping or through an API. For the majority of these paintings we also gathered the following metadata: title of the work, estimated year of creation and name of the artist. 

\begin{figure}[h!]
    \begin{center}
    \includegraphics[width=\linewidth]{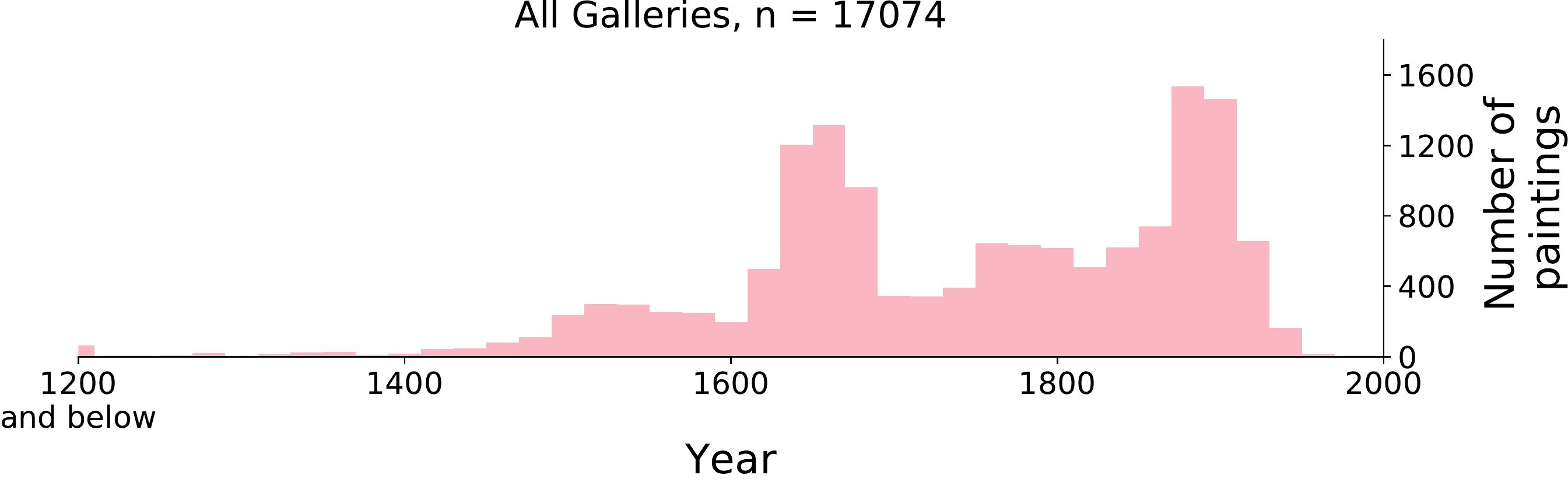}
    \caption{Histogram of the distribution of paintings over time.
Each bin equals 20 years.  }
    \label{year_dist_all}
    \end{center}
\end{figure}

For 92\% of boxes, we also have an estimate of the year of production. These estimates were made by the galleries from which the paintings were downloaded. The distribution of the year of production for all paintings are plotted in \fig{year_dist_all}, in bins of 20 years. Clearly, the data is not equally distributed over time. Museums, being cultural institutions, attempt to create a curated collection of art of cultural importance and not every year or time period is of equal cultural importance.

\begin{table}[h!]

    \centering
    \begin{tabular}{lll}
    Gallery Name	                & Country & Count        \\
    \hline
    The Rijksmuseum	                & Netherlands &	4672	\\
    The Metropolitan Museum of Art  & USA   &	3222         \\
    Nationalmuseum                  & Sweden &	3077        \\
    Cleveland Museum of Art         & USA &	2217   \\
    National Gallery of Art         & USA &	2132   \\
    Museo Nacional del Prado        & Spain &	2032   \\
    The Art Institute of Chicago    & USA &	936	    \\
    Mauritshuis                     & Netherlands &	638     \\
    J. Paul Getty Museum            & USA &   399     \\
    \end{tabular}
    \caption{List of galleries, the country in which the museum is located, and the number of paintings downloaded from that gallery.}
     \label{galleries}

\end{table}

\subsection*{Image-level coarse-grained material labels}
Next, we collected human annotations to identify material categories within paintings. We created a list of 15 material categories: animal, ceramic, fabric, sky, stone,
flora, food, gem, wood, skin, glass, ground, liquid, paper, and metal.
Our intention was to create a succinct list, that would nevertheless
allow the majority of stuff within a painting to be annotated, and was partially based on \cite{Sharan2009, Fleming2013}. These categories primarily represent prototypical physical materials, as well as less typical, overarching categories that contain multiple materials such as \textit{food} and \textit{animal}. Note that we added the material category of \textit{skin} directly, instead of a more overarching `human' category as one might expect considering \textit{food} and \textit{animal}. We made this choice because of the scientific interest in the artistic depiction \cite{lehmann2008fleshing}, perception \cite{Stephen2011, Matts2007}, and rendering of skin \cite{Igarashi2007, Jensen2001}

In one AMT task, participants would be presented with 40 paintings at
a time and one target material category.  In the task, participants
were asked if the painting depicted the target material (e.g.,
\emph{does this painting contain wood?}). They could reply \emph{`Yes,
the target material is depicted in this painting.'} by clicking the
painting and inversely, by not clicking the painting, participants
would reply with \emph{`No, the target material is not depicted in
this painting.'}. Each painting was presented to at least 5
participants for each of the 15 materials. If at least 80\% of the
responses per painting claimed that the material was depicted in the
painting, we would register that material as present for that
painting. In total, we collected 1,614,323 human responses in this
stage from 3,233 unique AMT users participating. 

\subsection*{Extreme click bounding boxes}
In the previous stage, paintings were registered to depict or not to
depict a material. However, that stage does not inform us (1) how
often the material is depicted,  nor (2) where the material(s) are within the painting. 

\begin{figure}[!ht]
    \centering
    \includegraphics[width=0.4\linewidth]{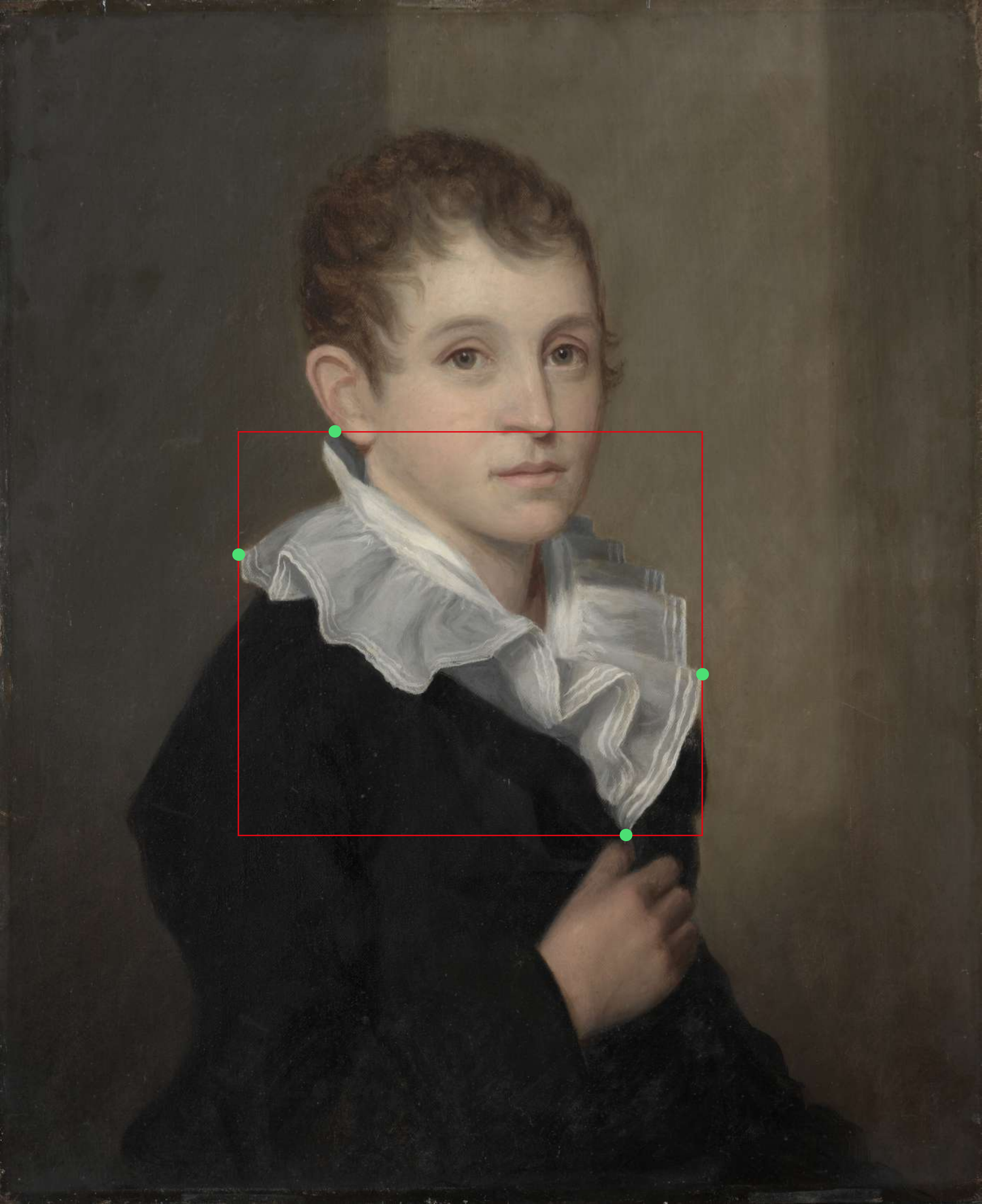}
    \caption{An example of four extreme clicks (marked in green) made by a user on a piece of fabric. These points correspond to the most left, most right, highest and lowest points on the annotated item. The red-line displays the resulting bounding box.}
    \label{extreme_clicks_example}
\end{figure}

We gathered this information on the basis of extreme click bounding
boxes. For extreme click bounding boxes, a participant is asked to
click on the 4 extreme positions of the material: the highest, lowest,
most left-, and most right-wards point \cite{Papadopoulos2017}. See \fig{extreme_clicks_example} 
for an example. In the task, participants were presented with paintings that depicted the target material and tasked to create up to 5 extreme click bounding boxes for the target material. 

To make bounding boxes within the task, the participants would use our
interface, which allows users to zoom in and out, and pan around the
image. The interface furthermore allowed participants to finely adjust
the exact location of the extreme points by dragging the points
around. Initially, the tasks were open to all AMT workers, but after
around 2000 bounding boxes were created by 114 AMT users, with manual
inspection, we found that the quality of bounding boxes varied greatly
between participants. Therefore, we restricted the work to a smaller
number of manually selected participants who were observed to create
good bounding boxes. After this restriction, new boxes were manually
inspected by the authors, and in a few cases additional participants
were restricted due to a deterioration of bounding box quality.
Simultaneously additional participants were granted access to our
tasks after passing (paid) qualification tasks.  As a result, the
number of manually selected participants varied between 10 and 20
participants. In total, \numberOfBoxes bounding boxes were created by participants.

\paragraph{Automatic bounding boxes.}
While we consider our dataset to be quite larger, it only covers a small but representative portion of art history. It might be required to access materials in paintings that are not part of our dataset. To allow for this, we have trained a FasterRCNN \cite{fasterrcnn} bounding box detector to localize and
label material boxes in unlabelled paintings. The model was finetuned from a model trained on COCO with the COCO hyperparameters from \cite{detectron2}. First we trained the detector on 90\%
of annotated paintings in the dataset. In section \ref{automatic_boxes_eval} below, we show our evaluation of the network, which was performed on the remaining 10\% of annotated paintings. While we created this network to be able to detect paintings outside our dataset, we decided to apply the network on our dataset in order to more densely annotate our paintings. Therefor, after the evaluation, we ran the detection network on the entire set of paintings, i.e., training and testing data, in an attempt to more exhaustively annotate materials within paintings. From the automatic detected bounding boxes we first removed all boxes that scored $<$50\%  confidence (as calculated by FasterRCNN). Next, we filtered out automatic boxes that were likely already identified by human annotators, be removing automatic bounding boxes that scored $\geq{50}$\% on intersection over union, i.e., automatic boxes that shared the majority of it's content with human boxes. This resulted in an additional \automaticnumberOfBoxes bounding boxes, all of which are also available on \dbwebsite.

\subsection*{Fine-grained labels}
In this step we supplemented the previously collected material labels
with fine-grained material labels (see \tabl{subsetlabels}). For
example, a bounding box labelled as \emph{fabric} could now be
labelled as \emph{silk, velvet, fur, etc.}.  We excluded bounding boxes that were too
small (e.g., \emph{width in pixels} $\times$ \emph{height in pixels}
$\leq{5000}$) and boxes that were labelled as \emph{sky, ground} or
\emph{skin} for which fine-grained categorizations were not annotated. 
We collected fine-grained labels for the remaining 150,693 bounding
boxes. Note that this only concerns the bounding boxes created by human annotations as no automatically detected boxes were assigned a fine-grained material label. For each of these 150,693 bounding boxes, we gathered responses
from 5 different participants. If the 5 responses reached an
agreement of at least 70\%, we would assign the agreed upon label to
the bounding box.  To guide the workers, we provide a textual
description for each fine-grained category for them to reference
during the task. We did not provide visual exemplars as we did not
want to bias the workers into template matching instead of relying on
their own perceptual understanding.

We found that it is non-trivial to define fine-grained labels in such
a way that they are concise, uniform and versatile (i.e., useful
across different scientific domains) while still being recognizable and/or categorizable by naive observers. We applied the following reasoning to select fine-grained labels: 
first, we tried to divide the materials into an exhaustive list with as few fine-grained labels as possible. For example, for `wood', each bounding box is either `processed wood' or `natural wood'.  
If an exhaustive list would become too long to be useful, we would include an `other' option. For example, for `glass' we hypothesized that the vast majority of bounding boxes would be captured by either `glass windows' or `glass containers'. However, to include all possible edge cases such as glass spectacles and glass eyeballs, we included the `other' option. 

A possible subset for `metal' we considered was `iron',`bronze',
`copper', `silver', `gold' , `other'. However, we feared that naive
participants would not be able to consistently categorize these
metals. An alternative would be to subcategorize on object-level, e.g. 'swords', 'nails', etc., but as we are interested in material categorization, we tried to avoid this as much as possible. Thus, for `metal', and for the same reason `ceramic', we required a different method. We chose to subcategorize on color, as often the color for these materials are tied to object identity. 

Participants are shown one bounding box at a time and are instructed to choose which of the fine-grained labels they considered most applicable. Additionally, they are able to select a `not target material' option. 

We collected over one million responses from 1114 participants. 
This resulted in a a total of 105,708 boxes
assigned with a fine-grained label. See \tabl{subsetlabels} for
the numbers per category. 

\section*{Results and applications}

We conducted a diverse set of experiments to demonstrate how our annotated art-perception dataset can drive research across perception, art history, and computer vision. First, we report simple dataset statistics. Next, we organized our findings under the proposed dataset usage taxonomy: perceptual applications, ecological applications and computer vision applications.

\subsection*{Dataset statistics}

The final dataset contains painterly depictions of materials, with a total of \numberOfPaintings paintings. Participants have created a total of \numberOfBoxes bounding boxes and we additionally detected \automaticnumberOfBoxes using a FasterRCNN. Each box has a coarse material label and 105,708 also have been assigned a fine-grained material label.  The total number of instances per material categories (coarse- and fine-grained) can be found in \tabl{subsetlabels}.   Further analysis of spatial distribution of categories, co-occurences, and other related statistics will be discussed in a following section in the context of visual ecology.

\tablefirsthead{
  \\
  \toprule 
  Coarse-grained & Fine-grained & \# Labels\\ 
  \midrule
}

\tabletail{
    \midrule
    \multicolumn{2}{l}{{Continued on next page}} \\
}

\tablehead{
    \multicolumn{2}{l}{{Continued from previous page}} \\
    \toprule 
    Coarse-grained & Fine-grained & \# Labels \\ 
    \midrule
}

\tablelasttail{
  \midrule
 }

\bottomcaption{The number of annotated bounding boxes for each coarse- and
fine-grained category. Note that not every bounding box is associated with a
fine-grained label since participants were not always able to arrive at a
consensus. See main text for details. \label{subsetlabels}}

\begin{supertabular}{p{0.3\linewidth} p{0.4\linewidth} p{0.15\linewidth}} 
animal          &                                     & 11606 \\
                & birds                               &  1822 \\
                & reptiles and amphibians             &   144 \\
                & fish and aquatic life               &   289 \\
                & mammals                             &  7752 \\
                & insects                             &   155 \\
                & other animals                       &    10 \\
ceramic         &                                     &  3641 \\
                & brown or red                        &  1088 \\
                & white                               &   381 \\
                & decorated                           &   289 \\
                & other ceramic                       &    14 \\
fabric         &                                     & 31557 \\
                & velvety                             &   261 \\
                & lace                                &   491 \\
                & silky/satiny                        &  1354 \\
                & cotton/wool-like                    &  5712 \\
                & brocade                             &    96 \\
                & fur                                 &    27 \\
                & other fabric                       &    12 \\
flora           &                                     & 26693 \\
                & trees                               & 12851 \\
                & vegetables                          &    96 \\
                & fruits                              &  1238 \\
                & flowers                             &  2515 \\
                & plants                              &  3699 \\
food            &                                     &  3690 \\
                & cheese                              &    11 \\
                & vegetables                          &   107 \\
                & fruits                              &  1536 \\
                & meat or poultry                     &   183 \\
                & bread                               &   127 \\
                & seafood                             &   183 \\
                & nuts                                &     8 \\
                & other                               &    14 \\
gem             &                                     & 10525 \\
                & pearls                              &   719 \\
                & gemstones                           &   715 \\
                & other gems                          &     1 \\
glass           &                                     &  5546 \\
                & glass window                        &  2243 \\
                & glass container                     &  1003 \\
                & other glass                         &   171 \\
ground          &                                     &  2552 \\
liquid          &                                     &  5737 \\
                & body of water                       &  4583 \\
                & liquid in container                 &   458 \\
                & other liquid                        &   172 \\
metal           &                                     & 27708 \\
                & colorless metal                     &  2933 \\
                & yellowish metal                     &  4435 \\
                & brownish or reddish metal           &   510 \\
                & multicolored or other colored metal &   215 \\
paper           &                                     &  3167 \\
                & paper book                          &  1380 \\
                & paper sheets                        &   585 \\
                & paper scrolls                       &   114 \\
                & other paper                         &    19 \\
skin            &                                     & 32323 \\
sky             &                                     & 12734 \\
stone           &                                     & 23157 \\
                & processed stone                     &  9226 \\
                & natural stone                       &  9429 \\
wood            &                                     & 26953 \\
                & processed wood                      & 12810 \\
                & natural wood                        & 10751 \\

\end{supertabular}

\subsection*{Perceptual applications}
We believe that the materials in this dataset can be useful as stimuli for perceptual experiments. We demonstrate this in this section by performing an annotation experiment to study perception-based recipes. 

\subsubsection*{Perception-based recipes in painterly depictions \label{sec:perceptual_shortcut}}
As previously argued, we believe that painterly techniques are a sort of perception-based recipe. Applying these recipes results in a stylized depiction that can trigger a robust perception of the distal world. Studying the image features in paintings can lead to an understanding of what cues the visual systems needs to trigger a robust perception. 

Here we explore a perceptual shortcut for the perceptions of glass by annotating highlights in paintings and comparing these with highlights in photos.
For input stimuli, we use bounding boxes from our dataset and photographs sourced from COCO
\cite{lin2014}. Participants for this study included 3 of the authors, and one
naive lab-member.

\paragraph{Stimuli.}
We used 110 images of drinking glasses. First, we selected all bounding boxes in the \textit{glass, liquid container} category in our dataset. From this set, we manually selected drinking glasses, since this category can also contain items such as glass flower vases. Next, we removed all glasses that were most occluded, were difficult to parse from the background - for example when multiple glasses were standing behind each other, and removed images smaller then 300x300 pixels. This resulted in a few hundred painted drinking glasses.

Next, we downloaded all images containing cups and wineglasses from the COCO \cite{lin2014} dataset, from which we removed all non-glass cups, occluded glasses, blurry glasses and glasses that only occupy a small portion of the image, and small images. This left us with 55 photos of glass cups and wineglass. Next, we randomly selected 55 segmentations from our painted glass collection. Each stimulus was presented in the task at 650 $\times$ 650 pixels, keeping aspect ratio intact. 

\begin{figure}[h!]
\includegraphics[width=\linewidth]{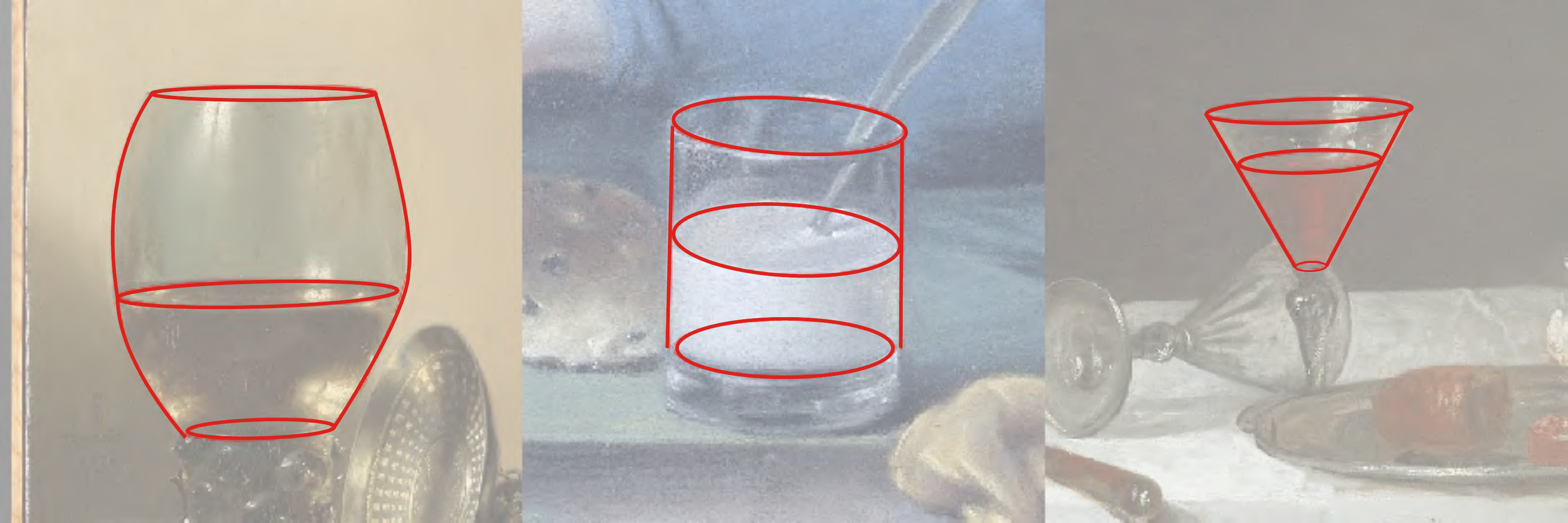}
\caption{Examples of the three glass shapes. From left to right: spherical, cylindrical and conical. The red  geometry annotations were manually created by the authors, and were used to standardize across glasses for the highlight analysis. }
\label{glass_catagories_horizontal}
\end{figure}

During this selection phase, we did not base our decision on the shape of the glass. After the experiment, as part of the analysis, we divided the glasses into three shapes, namely spherical, cylindrical, and conical glasses. See  \fig{glass_catagories_horizontal} for an example of each shape. 

\paragraph{Task.}

Participants annotated highlights on drinking glasses using an
annotation interface. In the annotation
interface, users would be presented with an image on which the
annotated geometry was visible. This made it clear which glass should
be annotated, in case multiple glasses were visible in the image. Users were instructed to instruct all visible highlights on that glass. Once
the user started annotating highlights, the geometry would no longer be visible. Annotations could be made by simply holding down the left-mouse button and drawing on top of the image. Once a highlight was annotated a user could mark it as finished and continue with the next highlight, and eventually move to the next image.
 
\paragraph{Results.}

\begin{figure}[h!]
    \includegraphics[width=\linewidth]{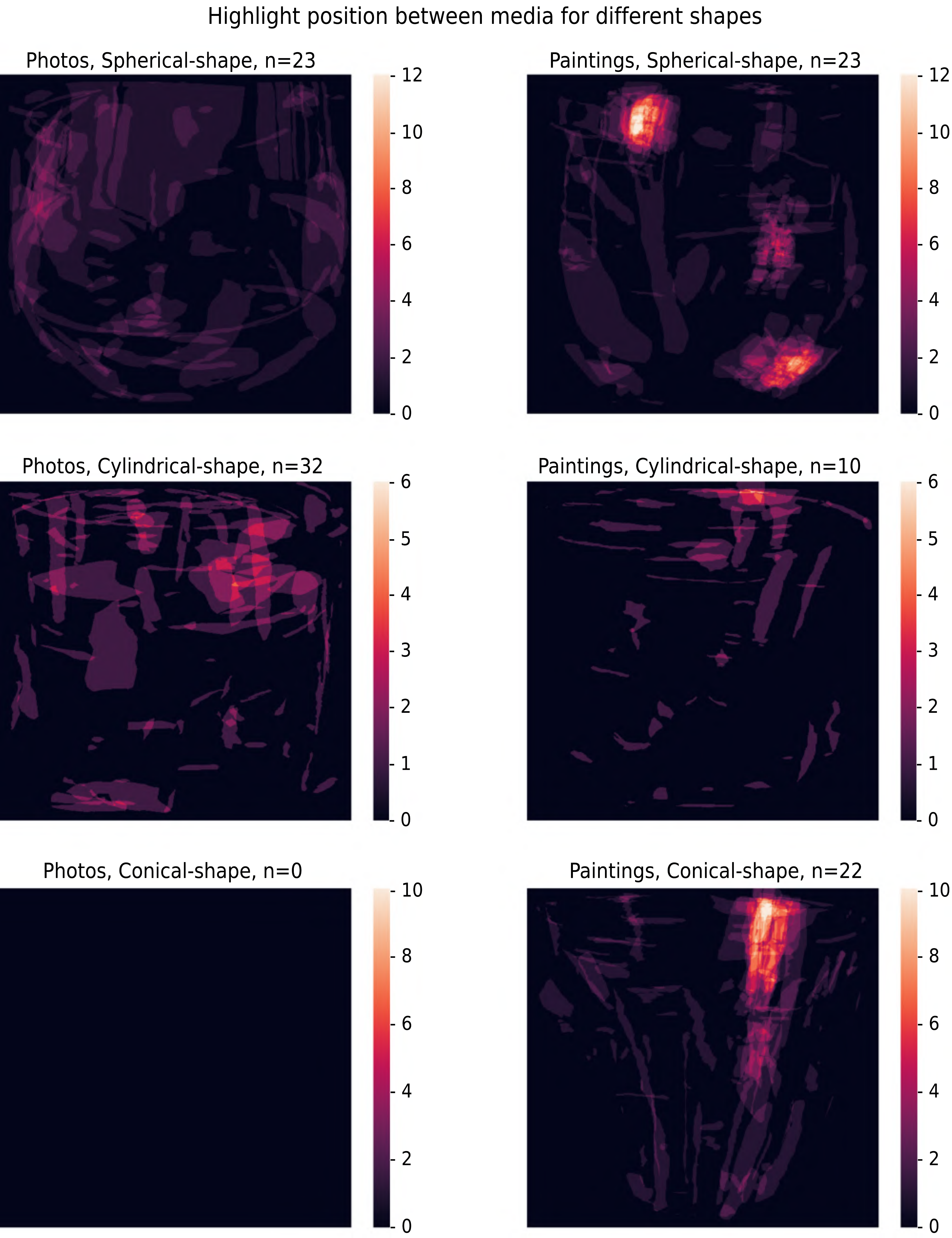}
    \caption{The overlaid highlights created by users, split on media and glass shape. In general, the photographic glass shapes display more variability and do not display a clear pattern. Note that for photos, no stimuli existed with a conical shape in our set which leads to a black image, since there were no highlight-annotations. On the right, for painted glasses, we see clear patterns in the placement of highlights for each glass shape. }
    \label{heatmaps_highlights_shape}
\end{figure}

To compare the highlights between photos and stimuli, we resized each glass to have the same maximum width and height, and then overlaid each stimuli on the center. When all 110 stimuli are overlaid (not
visualized) the resulting figure is appears noisy. However,
when we split the stimuli on media and shape, a clear pattern emerges for painted stimuli \fig{heatmaps_highlights_shape}. 
 
As can be seen, painters are more likely to depict highlights on glasses adhering to a stylized pattern, at least for spherical and conical glasses. This pattern of highlights is perceptually convincing, but is perhaps surprisingly uniform in comparison with the variation found within reality. Furthermore, we calculated the agreement between each pair of participants,as the ratio of pixels annotated by both participants (i.e., overlapping area) divided by the number of pixels that was an annotated by either participant (i.e., total area). Averaged across participants, the agreement on  paintings (0.33) was around 50\% higher relative to the average agreement between participants on photos (0.21).  This means that for our stimuli, highlights in paintings are less ambiguous when compared to photos. 

\subsection*{Ecological applications} 

The ecology displayed within paintings are representative of our visual
culture. Our dataset consists of paintings spanning 500+ years of
art history. This provides a unique opportunity to analyze a specific sub-domain of visual culture, i.e., that of paintings. We first analyze the presence of materials in paintings in the \textit{Material presence }section and in the next section we analyse this over time. In \textit{The spatial layout of materials} , we visualize the spatial distributions of materials in our dataset. In the last section, we analyze the automatically detected bounding boxes. 

\begin{figure}[h!]
    \begin{center}
    \includegraphics[width=1\linewidth]{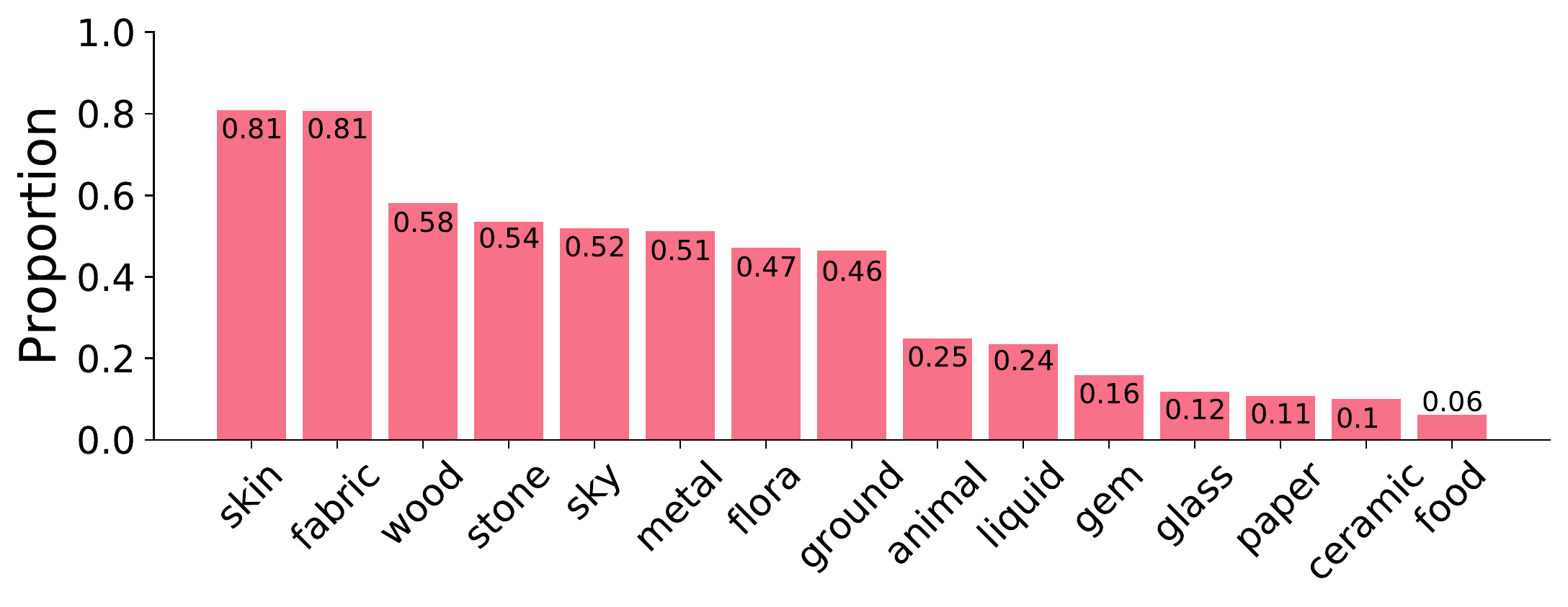}
    \caption{The proportion of paintings in our dataset that depict at least one instance of each material.}
    \label{material_presence}
    \end{center}
\end{figure}

\paragraph{Material presence.}\label{presence_section}
Within the \numberOfPaintings paintings, participants exhaustively
identified the presence of \numberOfMaterialInstances instances of 15 coarse materials. In other words, for each painting, participants indicated if each material is or is not present. The distribution of unique materials
per painting is normally distributed with an average of 5.7 unique
coarse materials present per painting (std = 2.8 materials). The most
frequent materials are \emph{skin} and \emph{fabric}. The least
frequent are ceramics and food. The relative frequency of each coarse material
is presented in \fig{material_presence}.  We did not exhaustively identify fine-grained materials within paintings, so we will not report those statistics here.  

\begin{figure}[h!]
    \begin{center}
    \includegraphics[width=1\linewidth]{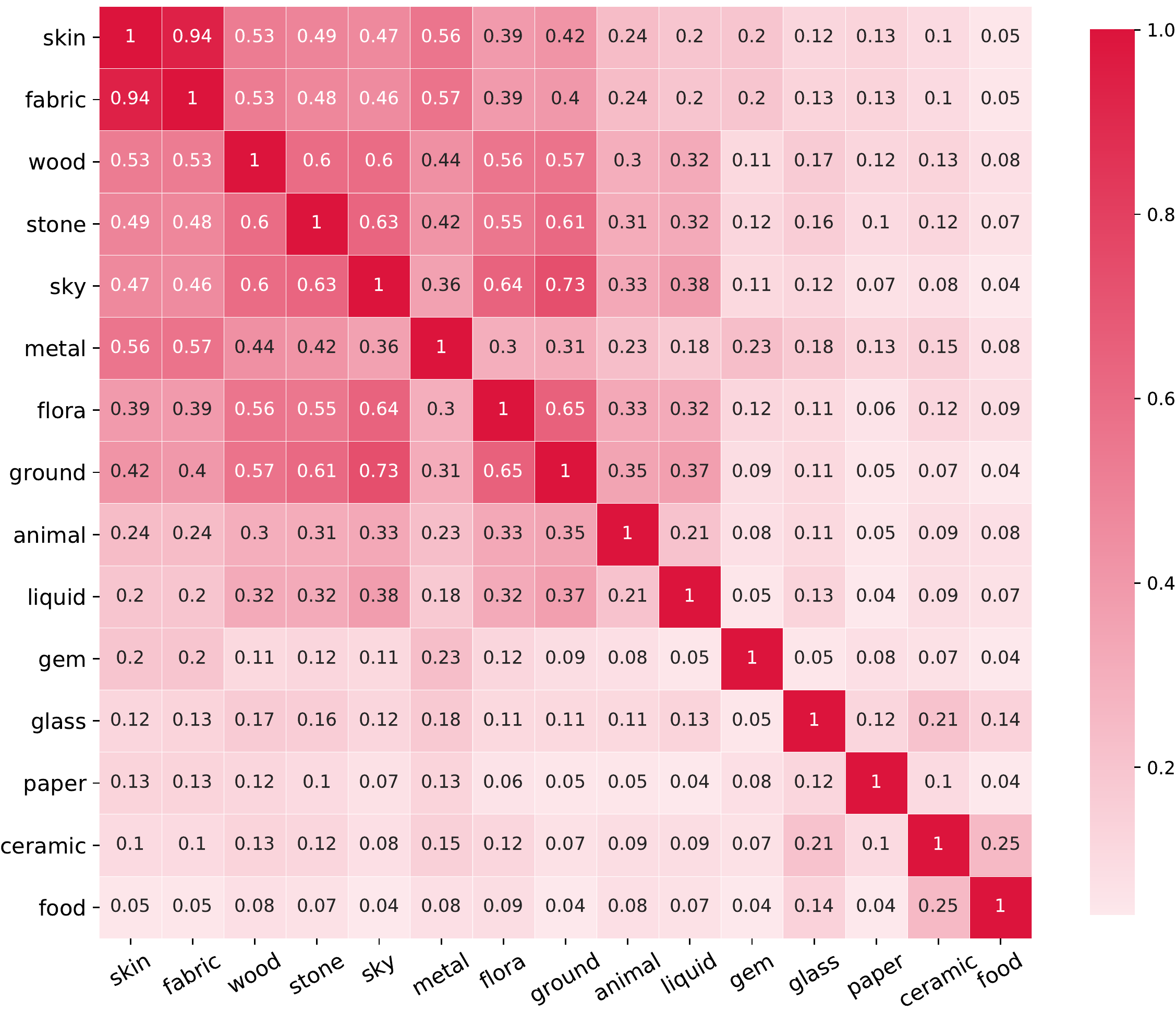}
    \caption{Co-occurrence matrix. Each cell equals the number of paintings where both materials are present divided by the number of paintings where one or the other material is present. }
    \label{cooccurrence}
    \end{center}
\end{figure}

Based on prior knowledge of natural ecology, one might assume that
some materials, such as \emph{skin} and \emph{fabric} might often be
depicted together in paintings. To quantify the extent to which
materials are depicted together, we create a
co-occurrence matrix presented in \fig{cooccurrence},  where each cell is the co-occurrence for each pair of materials as the number of paintings where both materials are present, divided by the number of paintings where either (but not both) materials are present. We can see for example, that if \emph{skin} is depicted, there is a 94\% change to also find \emph{fabric} in the same painting. 

\begin{figure}[h!]
    \begin{center}
    \includegraphics[width=1\linewidth]{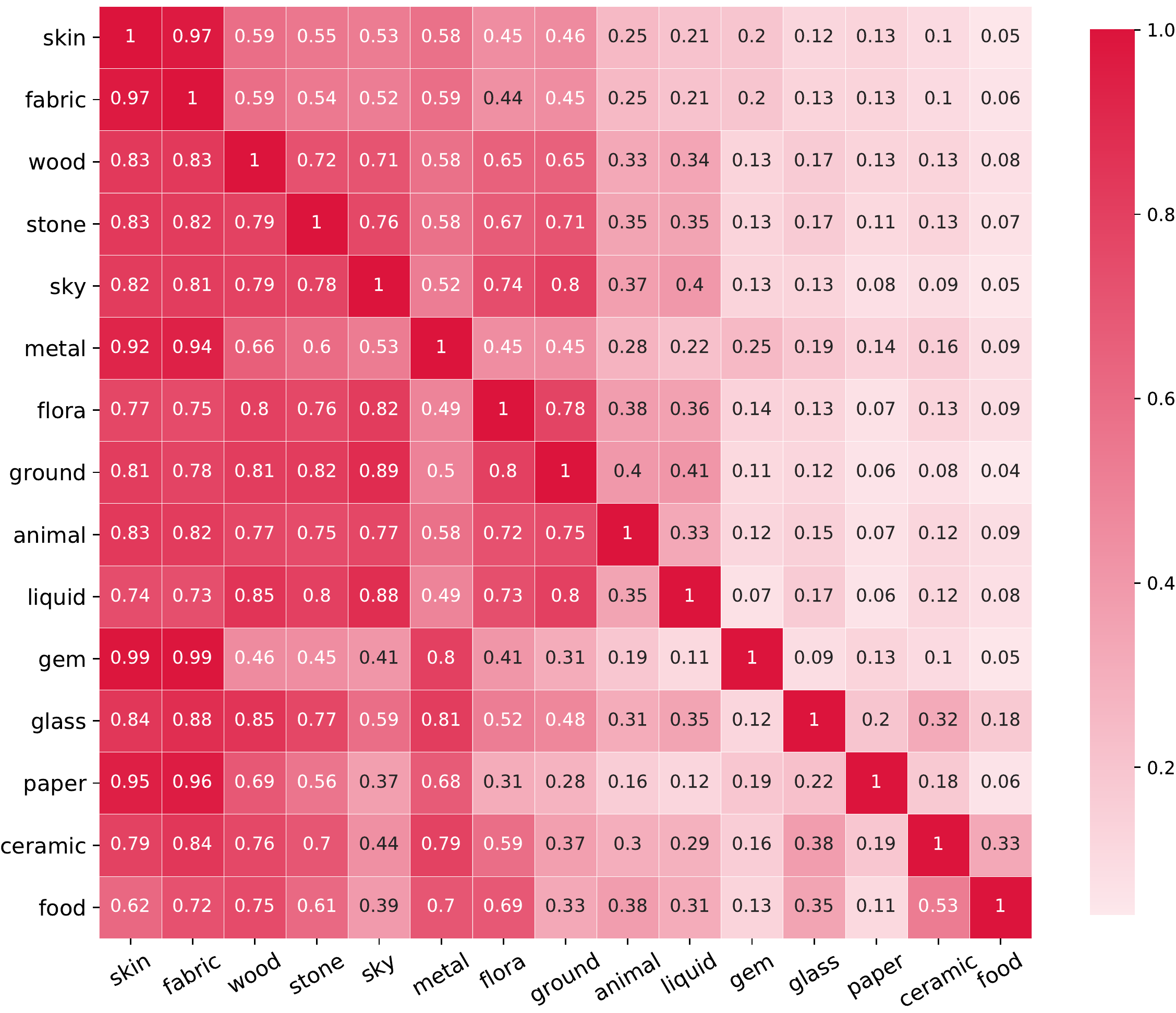}
    \caption{Likelihood matrix. This matrix visualizes the influence a
material has on the likelihood of finding another material within the
same painting, i.e., if one material on the y-axis is present, then how does this impact the presence of other materials on the x-axis? Calculated as the number of paintings where both materials are present, divided by the number of paintings that contain only one of the materials. }
    \label{if-then}
    \end{center}
\end{figure}

Furthermore, one might expect that the presence of one material can have an
influence on another material. For example, one might expect that \emph{gem}
might almost always be depicted with \emph{skin}, but that \emph{skin} is only
sometimes depicted with \emph{gem}. To quantify these relations, we calculated
the occurrence of a material given that another material is present. We
visualize this in \fig{if-then}. Here we see that \emph{if gem} is present,
\emph{then skin} is found in 99\% of the paintings, but that \emph{if skin} is
present, \emph{then gem} is found in only 20\% of the paintings. The same
relationship is true for gem and fabric. This implies that gems are almost
always depicted with human figures, however that human figures are not always
shown with gems. Another example, when liquid is present, in 85\% of the
paintings, wood is also present. One might be reminded of typical naval scenes,
or landscapes with forests and rivers. Inversely, when wood is present, only
34\% of the paintings depict liquid. For \emph{food} and \emph{ceramics}, two
materials which are present in less then 10\% of paintings, we see that if
\emph{food} is present, \emph{ceramics} has a 53\% change to be present as well,
but the inverse is only 33\%. This implies that food is served in, or with,
ceramic containers half of the time, but that this is only 1/3rd of what
ceramics is used for.

\paragraph{Material presence over time.}\label{time_section}

We have previously shown the distributions of materials in paintings in \fig{material_presence}.  When we created similar distributions (not visualised) for temporal cross-sections, for example for a single century, we found that these distributions were remarkably similar to the average distribution in \fig{material_presence}. We used t-tests, to see if the distribution for any century was significantly different from the average distribution in \fig{material_presence} and found no significant effect. This means that despite the changes in stylistic and artistic techniques over time, the distribution of materials (such as in \fig{material_presence} remained remarkably stable over time for the period covered in our dataset. 

\paragraph{The spatial layout of materials.}\label{spatial_section}
Paintings are carefully constructed scenes and it follows that a painter would carefully choose the location at which to depict a material. For example, \cite{Tyler1998}  reported a strong spatial convention to center one eye within portraits. With the knowledge that spatial conventions exists within paintings, it makes sense to assume these might extend to materials. The average spatial location and extent of materials is visualized by taking the (normalized) location of each bounding box for a specific material and subsequently plotting each box as a semi-transparent rectangle. The result is a  material heatmap, where the brightness of any pixel indicates the likelihood to find a material at that pixel. In this section, we limit the material heatmaps to only include the bounding boxes created by human annotators. In the next section, we visualize the material heatmaps for automated boxes too. 

\begin{figure*}[]
    \includegraphics[width=\linewidth]{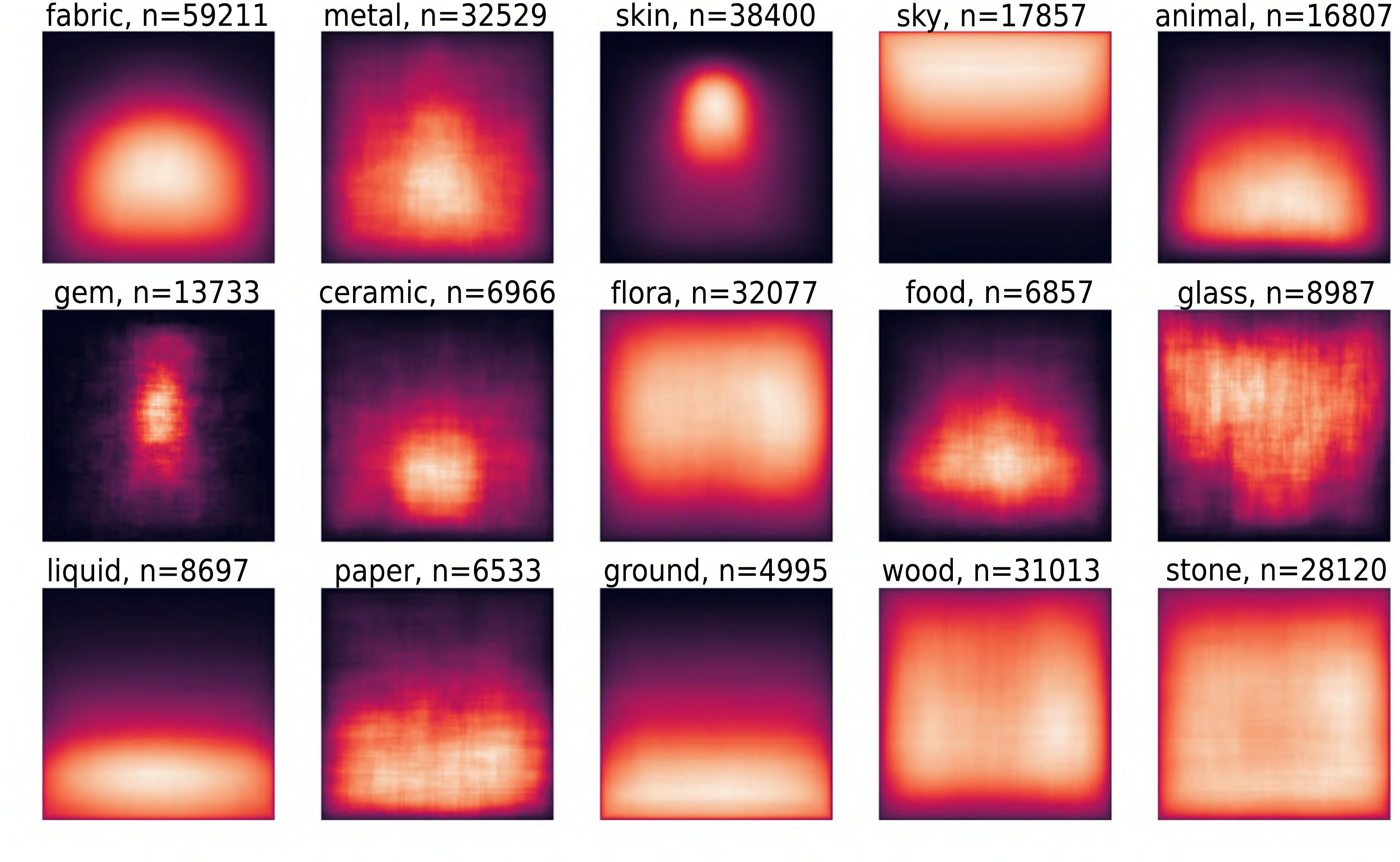}
    \caption{Material heatmaps, which illustrate the likelihood at any
given pixel to find the target material at that pixel. Brighter
colors indicate higher likelihoods.}     \label{heatmaps_2}
\end{figure*}

\begin{figure}[!h]
    \centering
    \includegraphics[width=0.6\linewidth]{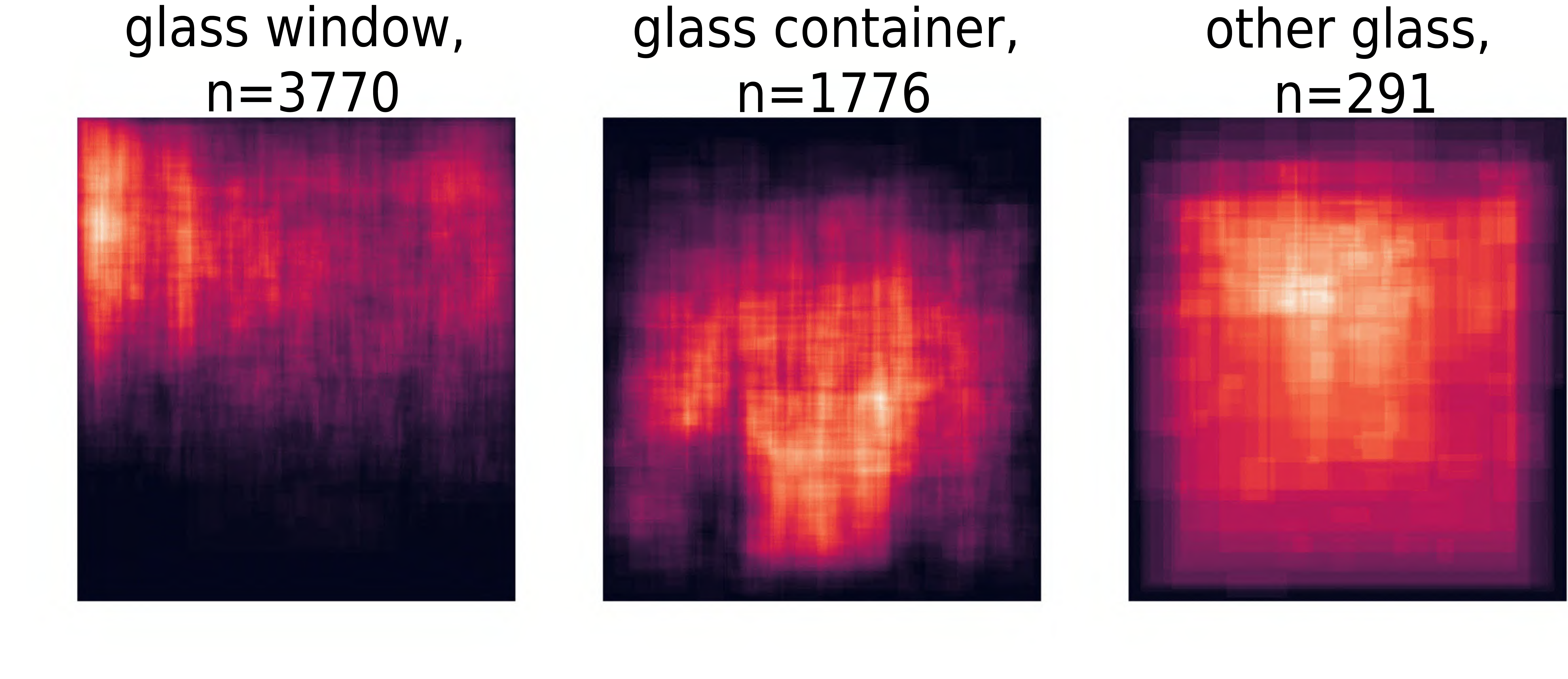}
    \caption{Material heatmaps for glass sub categories. For glass windows, it is interesting to see the clustering in the top-left corner, which is in agreement with the artistic convention of having light come from the top left.}
    \label{fig:heatmap_glass}
\end{figure}

Material heatmaps for the 15 coarse materials are shown in
\fig{heatmaps_2}. The expected finding that \emph{sky} and
\emph{ground} are spatially high and low within images serves as a
simple validation or sanity-check of the data. It is interesting to see how
\emph{skin} and \emph{gem} are both vertically centered within the
canvas. It appears to suggests a face, with necklaces and jewelry
adorning the figure. In general, each material heatmap appears to be
roughly vertically symmetric. For \emph{glass}, there does however
appear to be a minor shift towards the top-left. This might be related
to an artistic convention, namely that light in paintings usually
comes from a top-left window \cite{Carbon2018}. When we look at the
heatmaps for the sub-categories for glass in \fig{fig:heatmap_glass}, we see that it is indeed glass windows that show the strongest top-left bias.

\paragraph*{Automatically detected bounding boxes.}\label{automatic_boxes_eval}

Besides the bounding boxes created by humans, we also trained a FasterRCNN network to automatically detect bounding boxes with 90\% of the data as training data. On the remaining unseen 10\% of paintings, the network detected 90,169 bounding boxes. We removed those with a confidence score below 50\%, which resulted in 24,566 remaining bounding boxes. In the section below, all references to the automated bounding boxes refer to these 24,566 bounding boxes.

\begin{figure}[ht!]
    \centering
    \includegraphics[width=\linewidth]{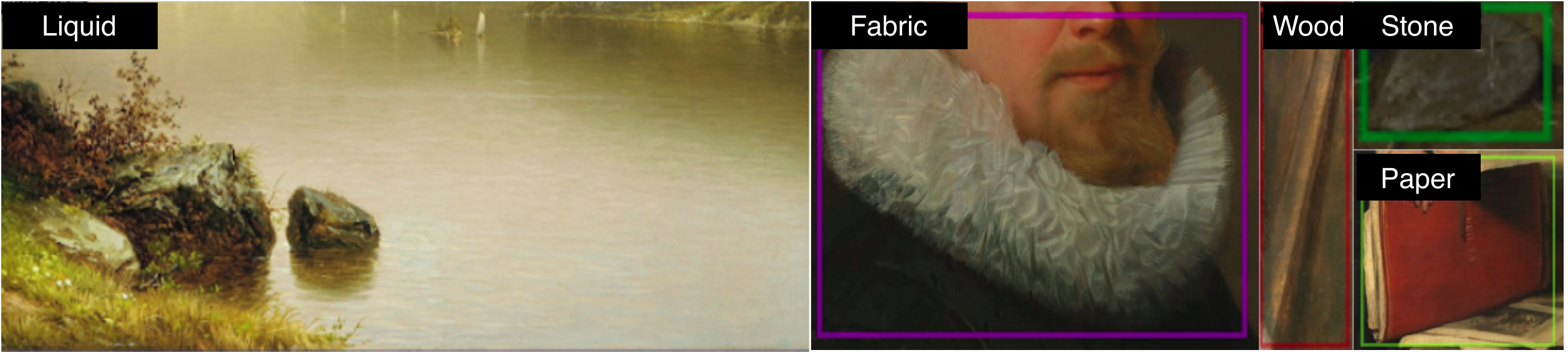}
    \vspace{-5mm}

    \caption{Examples of detected materials in unlabeled paintings.
    Automatically detecting materials can be useful for content retrieval for
    digital art history and for filtering online galleries by viewer interests.}

    \label{fig:rcnn}
\end{figure}

\begin{figure}[ht!]
    \centering
    \includegraphics[width=\linewidth]{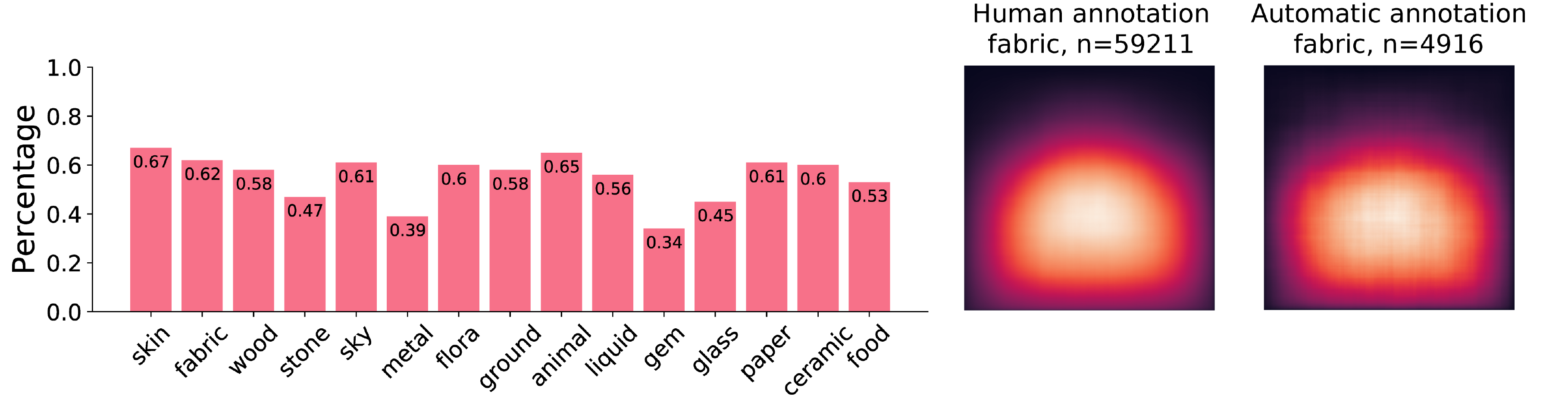}
    \vspace{-5mm}

    \caption{In the bar graph, the accuracy for automatically detected bounding boxes is displayed in the same order as in \fig{material_presence}. The values were derived from human quality votes. On the right, we compare the material heatmaps for fabric between the automated and the human annotation bounding boxes. }

    \label{automated_bounding_box_validation}
\end{figure}

A qualitative sample of detected bounding boxes is given in \fig{fig:rcnn}. Our human bounding boxes are non-spatially exhaustive in nature meaning that not every possible material has been annotated. As a result,  the automatically created bounding boxes can not always be matched against our human annotations and thus we can not use this to evaluate their quality. In order to validate the automatic bounding box detection, we performed a simple user study to get an estimate of the accuracy per material class, which is visualized in \fig{automated_bounding_box_validation}. In the user study, a total of 50 AMT participants judged a random sample of 1500 bounding boxes. The bounding boxes were divided into 10 sets of 150 stimuli, each set contain 10 boxes per course material class.  Each individual participant only saw one set, and each set was seen by 5 unique participants. The order of stimuli was randomized between sets and participants. A participant was instructed to rate each stimuli was either a \textit{good} or a \textit{bad} bounding box.  This leads to a total of 7500 votes, 500 per material classes. The ratio of good to bad votes per material classes can serve as a measure of accuracy, which  has been visualized in \fig{automated_bounding_box_validation}. 

As a result of the user study, we found a mean average accuracy of 0.55.  While not high, these results are somewhat interesting in that they show that a FasterRCNN model is capable of detecting materials in paintings, without any changes to the network architecture or training hyperparameters. It is certainly promising to see that an algorithm designed for object localization in natural images can be readily applied to material localization in paintings. Likely, the accuracy could be further improved by finetuning the network which we have not done in this paper. 

It is interesting to note that the spatial distribution of automatically detected bounding boxes looks very similar to the spatial distribution of the human annotated bounding boxes. We have visualized the material heatmap for one material, fabric, for the automated bounding boxes to show the similarity with the material heatmap for the same material created from human annotation bounding boxes. This has been visualized in the right side of \fig{automated_bounding_box_validation}

\subsection*{Computer vision applications}

In this section, we will first apply existing segmentation tools designed for natural photographs to extract polygon segmentations. Next, we perform an experiment to demonstrate the utility of paintings for automated material classification.

\subsubsection*{Extracting polygon segmentations} 

A natural extension of material bounding boxes is material segments
\cite{bell13opensurfaces, minc, caesar2018coco}.  Polygon
segmentations are useful for reasoning about boundary relationships between
different semantic regions of an image, as well as the shape of the regions
themselves.  However, annotating segmentations is expensive and many modern
datasets rely on expensive manual annotation methods
\cite{bell13opensurfaces,lin2014,ade20k,caesar2018coco,cityscapes}. Recent work has focused
on more cost effective annotation methods (e.g.
\cite{lin2019block,maninis2018deep,benenson2019large,ling2019fast}). One broad
family of methods to relax the difficulty of annotating polygon segmentations is
through the use of interactive segmentation methods that transform sparse user
inputs into a full polygon masks.

\begin{figure}[ht!]
    \centering
    \includegraphics[width=\linewidth]{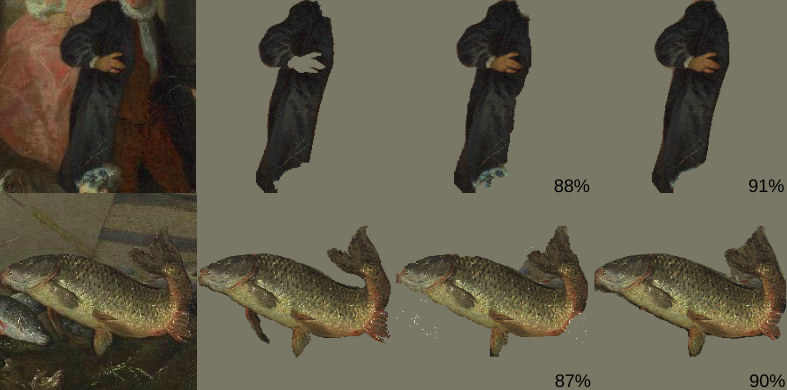}
    \caption{Segmentation visualizations. Left to right: Original Image, Ground Truth Segment, Grabcut
    Extr Segment, DEXTR COCO Segment. Both Grabcut and DEXTR use extreme points
    as input. For evaluation, the extreme points are generated synthetically
    from the ground truth segments. The IOU for each segmentation is shown in
    the bottom right corner.
    }
    \label{fig:segments}
\end{figure}

For this dataset, we apply interactive segmentation with the crowdsourced extreme
clicks as input. To evaluate quality, we compared against ~4.5k high-quality
human annotated segmentations from \cite{van2020painterly}, which were sourced
from the same set of paintings.  We find that both image-based approaches like
GrabCut (GC) \cite{grabcut} and modern deep learning approaches such as DEXTR
\cite{maninis2018deep} perform well.  Surprisingly, DEXTR transfers quite well to
paintings despite being trained only on natural photographs of objects. The
performance is summarized in \tabl{table:segments}. The performance is summarized using the standard intersection over union (IOU) metric. IOU is computed as the intersection between a predicted segment and the ground truth segment divided by the union of both segments. IOU is computed for each class, and mIOU is the mean IOU over all of the classes. Samples are
visualized in \fig{fig:segments}.  Segments produced by these methods from
our crowdsourced extreme points will be released with the dataset.

\begin{table}[ht!] 
  \centering 
  mIOU (\%) \\
  \begin{tabularx}{0.575\linewidth}{|c|c|c|c|c|}
  \hline
  \makecell[c]{\!\!Grabcut\!\!\\\!\!Rectangle\!\!} &
  \makecell[c]{\!\!Grabcut\!\!\\\!\!Extr\!\!} &
  \makecell[c]{\!\!DEXTR\!\!\\\!\!Pascal-SBD\!\!} &
  \makecell[c]{\!\!DEXTR\!\!\\\!\!COCO\!\!}
  &
  \makecell[c]{\!\!DEXTR\!\!\\\!\!Finetune\!\!}\\ 
  \hline
    44.1 & 72.4 & 74.3 & 76.4 & 78.4 \\ 
  \hline
  \end{tabularx}
  \vspace{2mm} \\
    DEXTR Finetune IOU By Class (\%) \\
  \begin{tabularx}{0.545\linewidth}{|c|c|c|c|c|} 
    \hline
    Animal &  Ceramic & Fabric & Flora & Food   \\ 
    \hline
    76.9 & 86.8 & 79.1 & 77.0 & 87.5 \\
    \hline
    \hline
    Gem & Glass & Ground & Liquid & Metal   \\
    \hline
    74.4 & 83.2 & 69.6 & 73.0 & 75.5 \\
    \hline
    \hline
    Paper & Skin & Sky & Stone & Wood \\
    \hline
    86.1 & 78.9 & 78.5 & 81.7 & 67.4 \\
    \hline
  \end{tabularx}
  \caption{Segmentations from extreme clicks.  Grabcut \cite{grabcut}
rectangles use bounding-box only initialization as a reference baseline. Grabcut Extr is
  based on the improved GC initialization from \cite{Papadopoulos2017} with
  small modifications: (a) we compute the minimum cost boundary with the cost as
  the negative log probability of a pixel belonging to an edge; (b) in addition
  to clamping the morphological skeleton, we also clamp the extreme points
  centroid as well as the extreme points; (c) we compute the GC directly on the
  RGB image. DEXTR \cite{maninis2018deep} Pascal-SBD and COCO are pretrained DEXTR
  ResNet101 models on the respective datasets. Note that Pascal-SBD and COCO are
  natural image datasets of objects, but DEXTR transfers surprisingly well
  across both visual domains (paintings vs. photos) and annotation categories
  (materials vs. objects).} 
  \label{table:segments}
\end{table}

\subsubsection*{Learning Robust Cues for Finegrained Fabric Classification}

The task of distinguishing between images of different semantic content is a
standard recognition task for computer vision systems. Increasing attention is
being given to ``fine-grained" classification where a model is tasked with
distinguishing images of the same broad category (e.g., distinguishing different
species of birds or different types of flora
\cite{wei2019deep,cub,inaturalist}). Fine-grained classification is particularly
challenging for deep learning systems. Such a task depends on recognizing
specific attributes for each finegrained class; in comparison, classifiers can
perform well on coarse-grained classification by relying on context alone. We
hypothesize that the painted depictions of materials can be beneficial for this
task. Since some artistic depictions focus on salient cues for perception
through perceptual shortcuts, it is possible that a network trained on paintings is able to learn a more robust feature representation by focusing on
these cues.

\paragraph{Task.} We experimented with the task of classifying cotton/wool versus
silk/satin.  The latter can be recognized through local cues such as highlights
on the cloth; such cues are carefully placed by artists in paintings. To
understand whether artistic depictions of fabric allow a neural network to learn
better features for classification, we trained a model with either photographs or
paintings. High resolution photographs of cotton/wool and silk/satin fabric and
clothing (dresses, shirts) were downloaded and manually filtered from publicly
available photos licensed under the Creative Commons from Flickr. In total, we
downloaded roughly 1K photos. We sampled cotton/wool and silk/satin samples from
our dataset to form a corresponding dataset of 1K paintings. We analyzed the
robustness of the classifier trained on paintings versus the classifier trained
on photos in two experiments below. Taken together, our results provide evidence
that a classifier trained on paintings can be more robust than a classifier
trained on photographs. 

\paragraph{Generalizability of classifiers.} Does training with paintings
improve the generalizability of classifiers? To test cross-domain
generalization, we test the classifier on types of images that it has not seen
before. A classifier that has learned more robust features will perform better
on this task than one that has learned to classify images based on more spurious
correlations. We tested the trained classifiers on both photographs and paintings.

In Table \ref{table:fg_fabric}, the performance of the two classifiers are
summarized. We found that both classifiers perform similarly well on the domain
they are trained on. However, when the classifiers are tested on cross-domain
data, we found that the painting-trained classifier performs better than the
photo-trained classifier. This suggests that the classifier trained on
paintings has learned a more generalizable feature representation for this task.

\begin{figure}[ht!]
    \begin{center}
      \includegraphics[width=\linewidth]{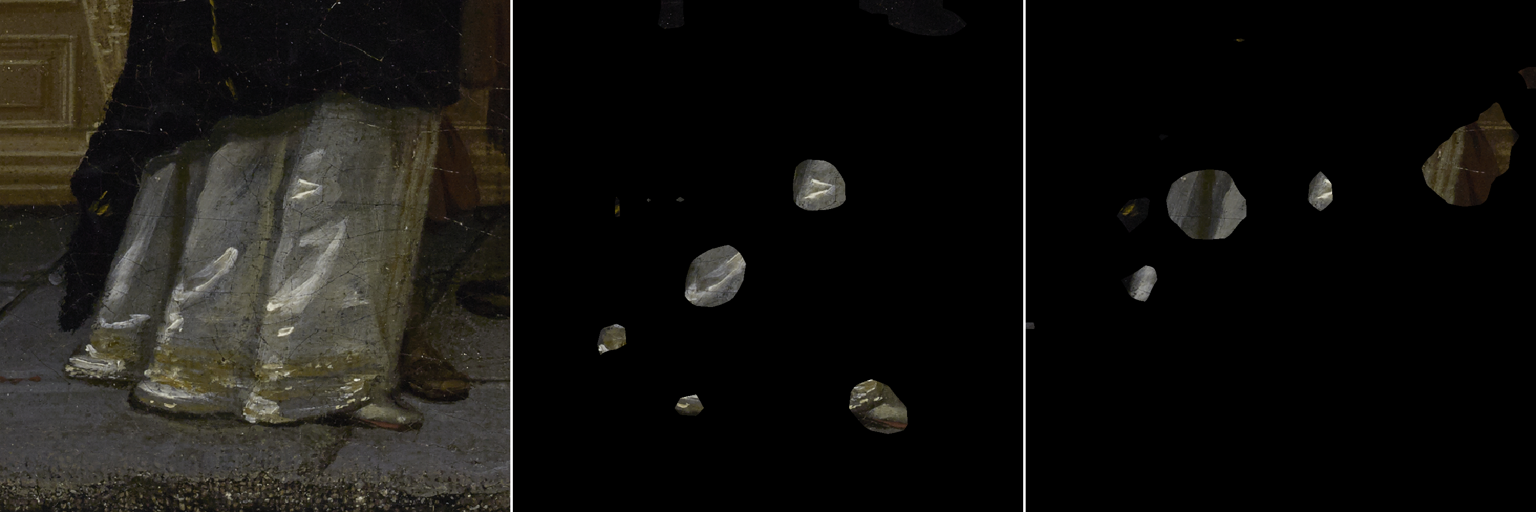}
      \caption{Visualization of cues used by classifiers. Left to right: Original Image, Masked Image (Painting Classifier), Masked Image (Photo Classifier). The unmasked regions
      represent evidence used by the classifiers for predicting ``silk/satin" in
      this particular image. See main text for details.}
    \label{fig:gradcam}
    \end{center}
\end{figure}

\paragraph{Human agreement with classifier cues.} How indicative are the cues
used by each classifier to humans? We produced evidence heatmaps with GradCAM
\cite{gradcam} from the feature maps in the network before the fully connected
classification layer.  We extracted high resolution feature maps from images of
size 1024 $\times$ 1024 (for a feature map of size 32 $\times$ 32). The heatmaps
produced by GradCAM show which regions of an image the classifier uses as
evidence for a specific class.  If the cues (i.e., \textit{evidence heatmaps},
such as in \fig{fig:gradcam}) are clearly interpretable, this would imply the
classifier has learned a good representation.  For both models, we computed
heatmaps for test images corresponding to their ground truth label. We conducted
a user study on Amazon Mechanical Turk to find which heatmaps are judged by human to be more informative. Users were shown images with regions corresponding to heatmap values
that are above 1.5 standard deviations above the mean. \fig{fig:gradcam}
illustrates an example. Users were instructed to 'select the image that contains the regions that look the most like <material>', where <material> was either cotton/wool or silk/satin. We collected responses from 85 participants, 57 of which were analyzed after quality control.  For quality control, we only kept results from participants who spent over 1 second on
average per trial.  

\begin{figure}[h!]
    \begin{center}
      \includegraphics[width=\linewidth]{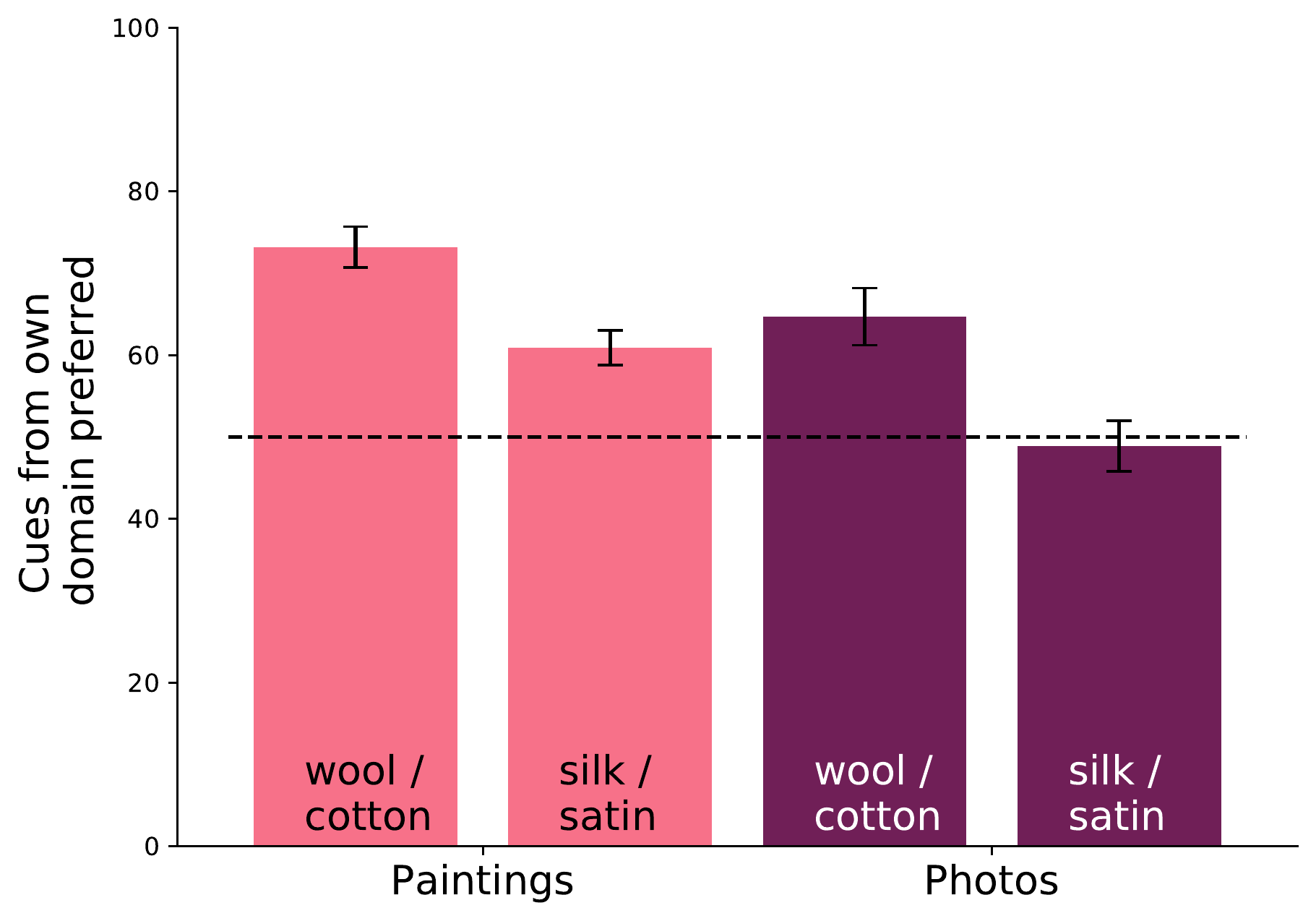}

      \caption{Human preference for classification cues used by each classifier.
      The y-axis represents how often humans prefer the cues from a classifier
      trained on the same domain as the test images. For example, the first bar
      indicates that in 73.2\% of the cases, humans preferred cues from the
      classifier trained on paintings when classifying wool/cotton paintings
      (and thus, the inverse, that in 26.8\% of the cases, humans preferred cues
      from the photo classifier.) Interestingly, note the last column -- humans
      equally prefer cues used by both classifiers for classifying silk/satin
      photos despite the painting classifier never seeing a photo during
      training. }

      \label{fig:fg_user_study_figure}
    \end{center}
\end{figure}

Overall, we found that the classifier trained on paintings uses evidence that is
better aligned with evidence preferred by humans
(\fig{fig:fg_user_study_figure}). Due to domain shifts when applying classifiers
to out-of-domain images, we would expect the cues selected by the painting
classifier to be preferable on paintings, and the cues selected by the photo
classifier to be preferable on photos.  Interestingly, this does not hold for
photos of satin/silk (see last column of \fig{fig:fg_user_study_figure}) --
found that users have no preference for the cues from either classifier, i.e.,
the cues from the painting classifier appears to be equally informative as the
cues from the photo classifier for categorizing silk/satin in photos.  This
suggests that either (a) the painting classifier has learned the ``key''
human-interpretable cues for recognizing satin/silk, or (b) that the photo
classifier has learned to classify satin/silk based on some spurious contextual
signals that are difficult to interpret by humans.  We asked users to elucidate
their reasoning when choosing which set of cues they preferred. In general,
users noted that they preferred the network which picks out regions containing
the target class. Therefore, it seems that the network trained on paintings has
learned better to distinguish fabric through the appearance of such fabrics
in the image over other contextual signals (see \fig{fig:gradcam}).

\begin{table}[h!]
    \centering
    \begin{tabularx}{0.7\linewidth}{|l||X|X|}
    \hline
      & \makecell[l]{Photo $\rightarrow$ Photo} & \makecell[l]{Painting$\rightarrow$ Painting} \\
    \hline
    \makecell[l]{MEAN F1 Score} & 79.6\% & 80.5\% \\
    \hline
    \hline
      & \makecell[l]{Photo $\rightarrow$ Painting} & \makecell[l]{Painting$\rightarrow$ Photo} \\
    \hline
    \makecell[l]{MEAN F1 Score} & 49.5\% & 57.8\% \\

    \hline 
    \end{tabularx}

    \caption{Classifier performance across domains. Classifiers are trained to distinguish
    cotton/wool from silk/satin.  The first column represents the classifier
    trained on photographs, and the second column represents the classifier trained
    on paintings. In the first row, the classifiers are tested on images of the
    same type they were trained on (i.e., trained and tested on photos, and
    trained and tested on paintings). In the second row, the classifiers are
    tested on the other medium, i.e., trained on photos and tested on
    paintings and vice versa. }

    \label{table:fg_fabric}
\end{table}

\section*{Discussion and conclusion} 

In this paper, we presented the Materials in Paintings (MIP) dataset -- a dataset of
painterly depictions of different materials throughout time. The dataset can be visited, browsed and downloaded at \dbwebsite. 

The MIP dataset consists of \numberOfPaintings high resolution images of
painting, in which we have annotated material information. Various datasets
exists that contain artworks, for example, the Painting-91 dataset from
\cite{khan2014painting} consists of around 4000 paintings from 91 artists and
was introduced for the purpose of categorization on style or artist. More
recently, Art500k was released, which contains more than 500k low resolution
artworks which were used to automatically learn to identify content and style
\cite{mao2017deepart} in paintings. Object recognition, while much more popular
on natural photographs, also has been performed on paintings such as in
\cite{crowley2014search} and \cite{crowley2014state}. It is worth mentioning the
WikiArt dataset, which is created by a non-profit organisation, with the
admirable goal ``to make world’s art accessible to anyone and anywhere" \cite{www.wikiart.org}. The WikiArt dataset has been widely used for a variety of scientific purposes \cite{bar2014classification, elgammal2018shape, saleh2015largescale, strezoski2017omniart, tan2017artgan}. 

What makes the MIP dataset unique is the availability of information on material depiction. To our knowledge, no datasets exist that provide annotations on material depictions in paintings, however a few datasets exists that provide material information for natural images. A notable example is OpenSurfaces \cite{bell13opensurfaces}, which contains around 70k crowd-sourced polygon segmentations of materials in photos. The Material In Context database improved on OpenSurfaces by providing 3 million samples across 23 material classes \cite{minc}. 

The availability of materials in paintings is beneficial to the research of material 
depiction and perception. In contemporary material perception (see \cite{fleming2017material} for a review) paintings are rarely used. A few noteworthy exceptions have already been mentioned within this paper, such as the work of Di Cicco \cite{di2020material} in which the authors studied the depiction of grapes in 17th century and used an explicit recipe written by a 17th century painter to recreate painterly depictions of grapes. An explicit, written-down recipe is not required for perceptual research as we have shown with our annotated highlights on glass. By only annotating pictorial cues the perception-based recipe can be revealed. Our findings raise an interesting question: what other perception-based recipes could generate insights into material perception? Does our finding of stylized highlights on glasses extend to other shiny materials? Within material perception, the perception of glossiness has received much attention \cite{ferwerda2001psychophysically, Chadwick2015, Wiebel2015, VanAssen2016}, and a better understanding of painterly depiction of glossy materials could be very beneficial. Of course, there is no reason why perception-based recipes should be limited to glass or glossiness, as painterly depictions are capable of conveying robust perceptions for many more perceptual attributes and materials \cite{van2020painterly} 

For art history,  the ability to easily access a large number of paintings that depict a
material might be interesting. Crowley and Zisserman \cite{Crowley14} pointed out that art historians often have the unenviable task of finding paintings for study manually. With the release of MIP, this task might now become slightly easier for art historians that study the artistic depiction of materials, such as for example stone \cite{Steinformen,dietrich1990rocks}. The fabric category, and it's fine-grained subclasses such as velvet, silk and lace could be used for the study of fashion and clothes in paintings in general \cite{hollander1993seeing, hollander2016fabric}, for paintings from a specific cultural context, such as Italian \cite{birbari1975dress}, English and French \cite{ribeiro1995art} or even for the clothes worn by specific artists \cite{de2005rembrandt}. The human body and it's skin, which clothing covers, is often studied within paintings \cite{hollander1993seeing, bol2012painting,lehmann2008fleshing}. For example, the Metropolitan Museum, published an essay on anatomy in the Renaissance, for which artworks depicting the human nude were used, many of which are incorporated in MIP \cite{bambach_2002}. In this work on anatomy, only items from the Metropolitan Museum were used but with the MIP this could be extended and compared to other museum collections. Furthermore, through for example the food and flora category, the MIP could give access to typical artistic scenes such as stillives \cite{grootenboer2006rhetoric, woodall2012laying} and floral scenes \cite{taylor1995dutch} respectively.

The usages of large sets of images has been a common practice for computer vision research. The usage of art datasets has been less common, but paintings have nevertheless been used in various ways. Models that learn to convert photographs into
painting-like or sketch-like images have been studied extensively for their
application as a tool for digital artists \cite{nstsurvey}.  Recent work has
shown that such neural style transfer algorithms can also produce images that
are useful for training robust neural networks \cite{SIN}. In a related paper, we more explicitly discuss specific applications of the MIP dataset for computer vision \cite{faper}. 
Similarly, other 
domains of computer vision research might benefit from painterly depictions.
The finding of our perception-based recipe for the stereotypical depiction of highlights on glasses 
could be useful for the generation of images. Current image generation algorithms
are capable of generating novel images, based on learned
statistics from a dataset \cite{van2016conditional,gregor2015draw}.
While a specific category of generated images, e.g., faces,
\cite{tolosana2020deepfakes} is rapidly becoming indistinguishable
from reality, a larger set of categories is still proving difficult to generate
due to the lack of sufficient training data. Moreover, it would be interesting
to see if applying explicit painterly techniques, e.g., perceptual
shortcuts, or stereotypical depictions could be leveraged in image generation. 
Perceptual shortcuts do not mimic the statistics of the real world, but instead capture image cues in a stylized depiction that explicitly trigger convincing human perceptions. Image generation algorithms that learn to use perceptual shortcuts might more efficiently capture image features that trigger perceptions.

Although the findings reported in this study are valuable for their own sake, 
we hope that the MIP dataset can support research in multiple disciplines, as well as promote multidisciplinary research. 
We have shown that depictions in paintings are not just of interest for
art history, but that they are also of fundamental interest for perception, as they can
illustrate what cues the visual system may use to construct a perception. Furthermore, paintings are explicitly created for human perception, which might be beneficial for algorithms trained on paintings. We have shown that computer vision algorithms trained on paintings appear to use cues more aligned with the human visual system, relative to algorithms trained on photos. The benefits of this might also extend to learning perceptually robust models for image synthesis.   

Our findings support our hope that the MIP dataset (freely accessible at \dbwebsite) will be a valuable addition to
the scientific community to drive interdisciplinary research in art history,
human perception, and computer vision. 

\section*{Acknowledgments}
We appreciate the work and feedback of AMT participants that participated in our user studies. We further wish to thank Yuguang Zhao for his help with the design of \dbwebsite. 

\bibliography{dataset_paper_submission}

\paragraph*{Paintings used.}
Below is a list of all paintings depicted within this paper. 

\begin{itemize}
    \vspace{-2mm}
    \item \textbf{ \fig{extreme_clicks_example}:} \textit{Samuel Barber Clark}, by \textit{James Frothingham}. 1810, Cleveland Museum Of Art
    
    \vspace{-2mm}
    \item \textbf{ \fig{glass_catagories_horizontal}, left:} \textit{Portret van een jongen, zittend in een raamnis en gekleed in een blauw jasje}, by \textit{Jean Augustin Daiwaille}. 1840, Het Rijksmusuem.
    
    \vspace{-2mm}
    \item \textbf{ \fig{glass_catagories_horizontal}, middle:} \textit{Still Life with Roemer, Silver Taza and Bread}, by \textit{ Pieter Claesz}, 1637, Museo Nacional del prado. 
    
    \vspace{-2mm}
    \item \textbf{ \fig{glass_catagories_horizontal}, right:} \textit{The White Tablecloth}, by \textit{Jean Baptiste Siméon Chardin}. 1731, The Art Institute of Chicago

    \vspace{-2mm}
    \item \textbf{ \fig{fig:rcnn}, liquid:} \textit{Lake George} , by \textit{John William Casilear}. 1857, The Metropolitan Museum of Art
    
    \vspace{-2mm}
    \item \textbf{ \fig{fig:rcnn}, fabric:} \textit{Man with a Celestial Globe} , by \textit{Nicolaes Eliasz Pickenoy}. 1624, The Metropolitan Museum of Art
    
    \vspace{-2mm}
    \item \textbf{ \fig{fig:rcnn}, wood:} \textit{The Monkey Sculptor}, by \textit{Teniers, David}. 1660, Museo Nacional del Prado
    
    \vspace{-2mm}
    \item \textbf{ \fig{fig:rcnn}, stone:} \textit{Thomas Howard, 2nd Earl of Arundel}, by \textit{Anthony van Dyck}. 1620, J. Paul Getty Museum
    
    \vspace{-2mm}
    \item \textbf{ \fig{fig:rcnn}, paper:} \textit{Portrait of Mr. Storer}, by \textit{Archer Shee, Sir Martin}. 1815, Museo Nacional del Prado
    
    \vspace{-2mm}
    \item \textbf{ \fig{fig:segments}, top:} \textit{Dance before a Fountain} , by \textit{Nicolas Lancret}. 1724, J. Paul Getty Museum
    
    \vspace{-2mm}
    \item \textbf{ \fig{fig:segments}, bottom:} \textit{Still life with fish}, by \textit{Pieter van Noort}. 1660, Het Rijksmusuem.

    \vspace{-2mm}
    \item \textbf{ \fig{fig:gradcam}:} \textit{Interior of the Laurenskerk at Rotterdam}, by \textit{Anthonie De Lorme, with figures attributed to Ludolf de Jongh}. 1662, J. Paul Getty Museum

\end{itemize}


\newpage
{
\bibliographystyle{plos2015}
}

\end{document}